\documentclass[traditabstract]{aa} 
\usepackage{graphicx,longtable,cancel}
\usepackage{txfonts,multirow}

\def\ltsima{$\; \buildrel < \over \sim \;$}
\def\lsim{\lower.5ex\hbox{\ltsima}}
\def\gtsima{$\; \buildrel > \over \sim \;$}
\def\gsim{\lower.5ex\hbox{\gtsima}}

\newcommand{\be}{\begin{equation}}
\newcommand{\en}{\end{equation}}

\def\cmdue {\rm \ cm^{-2}}

\def\deg {$^\circ$}

\usepackage{color}

\begin{document}

\title{Searching for {  narrow} absorption and {  emission} lines in XMM-{\it Newton} spectra of gamma-ray bursts}

\author{S. Campana\inst{1}, V. Braito\inst{1}, P. D'Avanzo\inst{1}, G. Ghirlanda\inst{1}, A. Melandri\inst{1}, A. Pescalli\inst{2,1}, \\
O. S. Salafia\inst{3,1}, R. Salvaterra\inst{4}, G. Tagliaferri\inst{1}, S. D. Vergani\inst{5}}
\institute{INAF-Osservatorio astronomico di Brera, Via Bianchi 46, I--23807, Merate (LC), Italy\\
\email{sergio.campana@brera.inaf.it}
\and
Dipartimento di Scienze, Universit\`a dell'Insubria, Via Valleggio 11, I--22100 Como, Italy
\and
Universit\`a di Milano-Bicocca, Piazza della Scienza 3, I--20126 Milano, Italy
\and
INAF-Istituto di Astrofisica Spaziale e Fisica Cosmica, Via Bassini, 15, I-20133, Milano, Italy
\and
GEPI - Observatoire de Paris Meudon, 5 Place Jules Jannsen, F-92195 Meudon, France
}

\date{}

\abstract
{
We  present  the results of a spectroscopic search for narrow emission and absorption features  in the X--ray spectra of  long 
gamma-ray burst (GRB) afterglows.  Using XMM-{\it Newton} data, both EPIC and RGS spectra,  of six bright (fluence $>10^{-7}$\, 
erg\, cm $^{-2}$) and relatively nearby ($z=0.54-1.41$) GRBs, we performed a blind search for emission or absorption 
lines that could be  related to a high cloud density or metal-rich gas in the environ close to the GRBs. 
We detected five  emission features  in four  of  the six GRBs with an overall statistical significance, assessed through Monte
Carlo simulations, of
$\lsim 3.0\,\sigma$.  Most of the lines are detected around  the  observed  energy of the oxygen 
edge at $\sim 0.5$ keV, suggesting that they are not related to the GRB environment but are most likely of Galactic origin.  
No significant absorption features were detected.
A spectral fitting  with a free Galactic column density ($N_H$)  testing different models for the Galactic absorption  confirms this 
origin because we found an indication of an excess of  Galactic $N_H$  in these  four GRBs with respect to the tabulated values. 
}

\authorrunning{Campana et al.} 

\maketitle

\section{Introduction}

Tracking the evolution of metals from star-sized scale to large-scale structures is a major step in understanding 
the evolution of the Universe. Metals are produced in stellar interiors and then ejected into their environments through 
supernova (SN) explosions and stellar winds, thus enriching the interstellar medium (ISM) of their galaxy. 
So far, the study of metal enrichment in galaxies at high redshifts has been carried out mostly by observing galaxies
along the line of sight of bright quasars (e.g. Prochaska et al. 2007). This technique is plagued by selection effects, 
e.g., radiation from quasars probes more probably halos of the intervening galaxies. Gamma-ray bursts (GRBs) are opening 
a  new window in understanding the history of metals and galaxy formation, in particular at high redshift. It is now 
recognised that long-duration GRBs are linked to collapse of massive stars, based on the association between 
(low--$z$) GRBs and (type Ic) core-collapse SNe (Woosley \& Bloom 2006). Many GRBs show 
intrinsic (i.e. in situ) X--ray absorption of the order of $10^{22}$ cm$^{-2}$ (Stratta et al. 2004; Campana et al. 2006a, 
2010, 2012, 2015; Watson et al. 2007, 2011; Behar et al. 2011, Eitan \& Behar 2013; Starling et al. 2013). 
Furthermore, the presence of absorption indicates a significant amount of metals because they are the only cause of 
photoelectric absorption of X--rays. Metallicities higher than those derived from damped 
Lyman-alpha (DLA) studies at the same redshift are also measured from optical studies of GRB. They indicate that 
the metal abundance can be as high as $10\%$ of the solar abundance up to $z\sim6$ (Savaglio 2012; Kawai et al. 2006; 
Campana et al. 2007).

Signatures of this material can be expected in the optical light curve of GRB afterglows 
as flux enhancements in their light curves (i.e. bumps) or, alternatively, as (variable) fine-structure transition lines in their spectra 
(Vreeswijk et al. 2007; D'Elia et al. 2009a, 2009b) as well as Ly$\alpha$ (Th\"one et al. 2011). 
Observations of these features can set important constraints on the density and distance of the absorbing material 
located either in the star-forming region within which the progenitor formed or in the circumstellar environment of the 
progenitor itself (Prochaska, Chen \& Bloom 2006; Dessauges-Zavadsky et al. 2006; D'Elia et al. 2009a, 2009b). 
Optical (high-resolution) studies can probe the circumstellar medium (CSM) only starting from $\gsim 100$ pc up to 
several kpc, but are not able to assess the innermost parts, probably because of the intense photoionisation field induced by the GRB.

\begin{table*}
\caption{XMM-{\it Newton} RGS observations of GRBs.}
\begin{center}
\begin{tabular}{ccccccc}
\hline
Name              &Redshift&Start time                & Time from GRB&Duration&Fluence $^*$    & Counts$^+$ (Net) \\ 
                         &               &                                 & (hr)                     & (ks)          &($10^{-7}$ erg cm$^{-2}$)& RGS1+2  \\
\hline
GRB 060729  &0.54       &2006-07-30 07:59& 12.8                  &    60.0     & 1.8                      & 13384 (10852) \\
GRB 061121  &1.31       &2006-11-21 21:42& 6.4                    &    38.4     & 1.0                      & 4847 (3763)\\
GRB 080411  &1.03       &2008-04-12 15:52& 18.6                  &    88.0     & 1.1                      & 10256 (6498)\\
GRB 090618  &0.54       &2009-06-18 13:47& 5.3                    &    20.5     & 1.9                       & 6406 (5371) \\
GRB 120711A&1.41       &2012-07-11 02:44& 21.8                  &    49.2     & 1.1                      &  5014 (3729)\\
\hline
GRB 111209  &0.68       &2011-12-09 22:29& 15.3                  &    53.6     & 0.5                      &  3142 (1901)\\
\hline
\end{tabular}
\end{center}
{\leftline{$^*$ Absorbed fluence in the 0.3--3 keV energy band for the pn instrument in units of $10^{-7}$ erg cm$^{-2}$.}}
{\leftline{$^+$ Total counts in the 0.45--1.8 keV energy band. In parenthesis the number of net source counts. }}

\end{table*}

Metal abundance derived from optical spectroscopy can be underestimated because some of the metals 
can be locked in dust grains, the composition of which can hardly be assessed through optical spectroscopy. This is not the case 
of X--ray measurements, for which X--rays are photoelectrically absorbed and scattered by dust grains as well. The soft X--ray 
band is best suited for metallicity diagnostics: absorption comes from the innermost shells (K) of elements from 
carbon (0.28 keV) to zinc (9.66 keV), with oxygen (0.54 keV) and iron (7.11 keV) being the most prominent. Iron L-shell 
edges are also detectable. In the X--ray domain, however, relatively little progress has been achieved in the study of GRBs. 
A transient feature consistent with being a Fe absorption has been reported for GRB 990705 and GRB 011211 
(Amati et al. 2000; Frontera et al. 2004) in the prompt emission phase. The statistical significance of the absorption features 
has been evaluated in detail only for GRB 011211 through parametric and non-parametric tests ($2.8-3.1\,\sigma$).

A decreasing total absorbing column density has been reported in one of the farthest GRB050904 (at $z = 6.3$), and 
this has been modelled, together with optical data, to reveal the high metallicity of the circumburst medium (Watson et al. 2006; 
Bo\"er et al. 2006; Campana et al. 2007, see also Butler \& Kocevski 2007). 
In a detailed study, Campana et al. (2007) placed the absorbing region at $\sim 10$ pc from the GRB site, testifying that X--ray studies can
probe the innermost region around GRBs. None of these studies were able to shed light on the chemical properties 
of the progenitor, however, because the relatively poor photon statistics did not permit a more detailed analysis.

\begin{figure*}
\hbox{
\includegraphics[width=6cm,angle=-90]{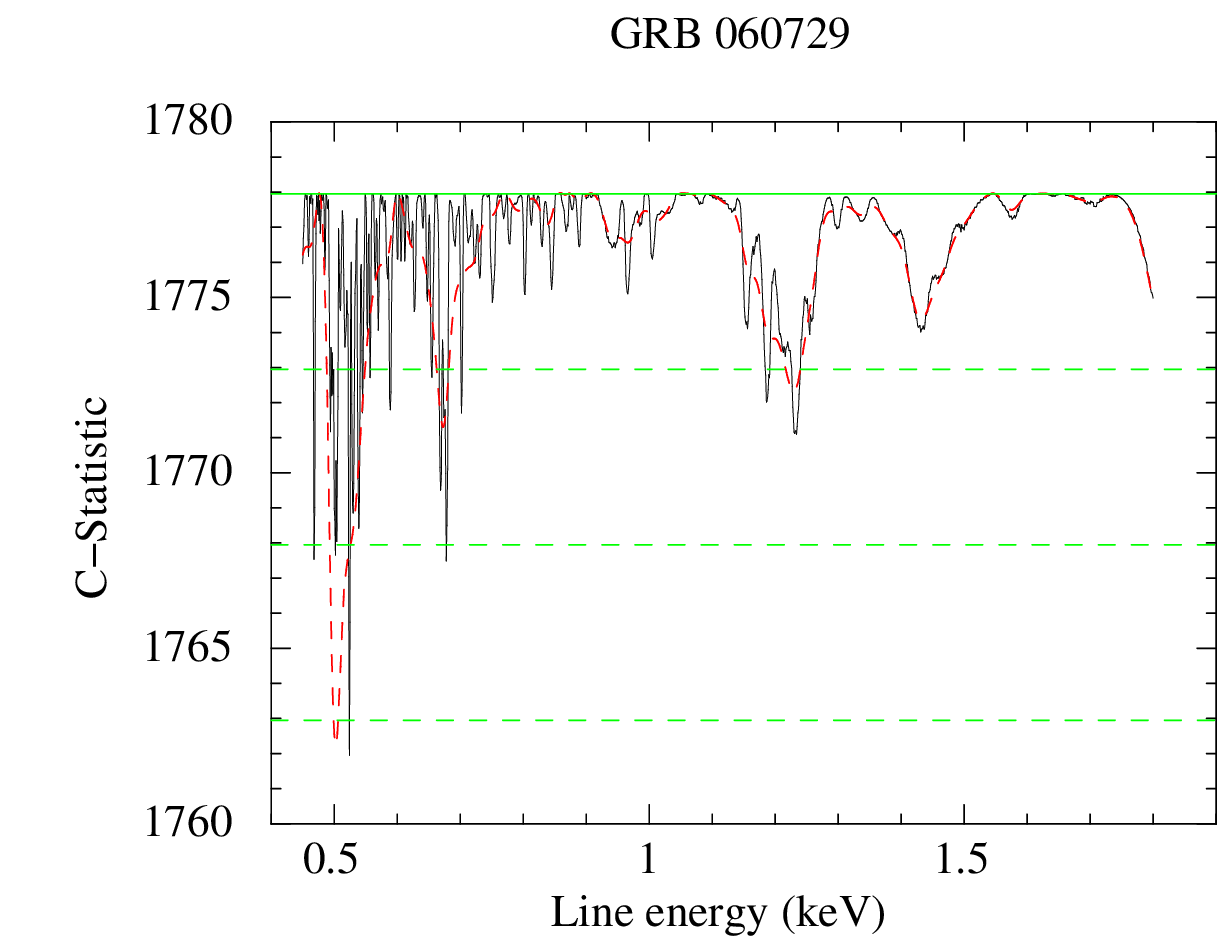}
\hspace{0.5cm}
\includegraphics[width=6cm,angle=-90]{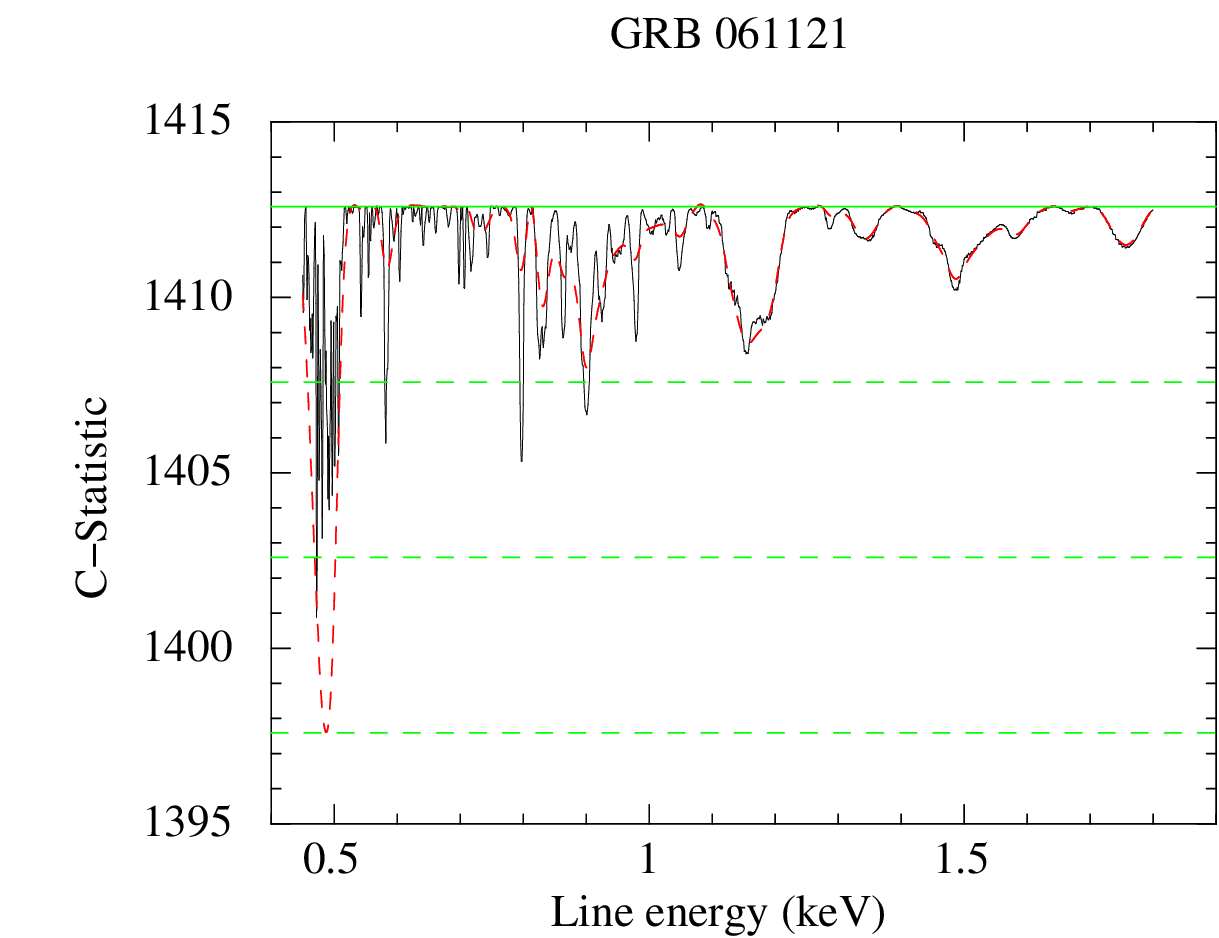}
}
\hbox{
\includegraphics[width=6cm,angle=-90]{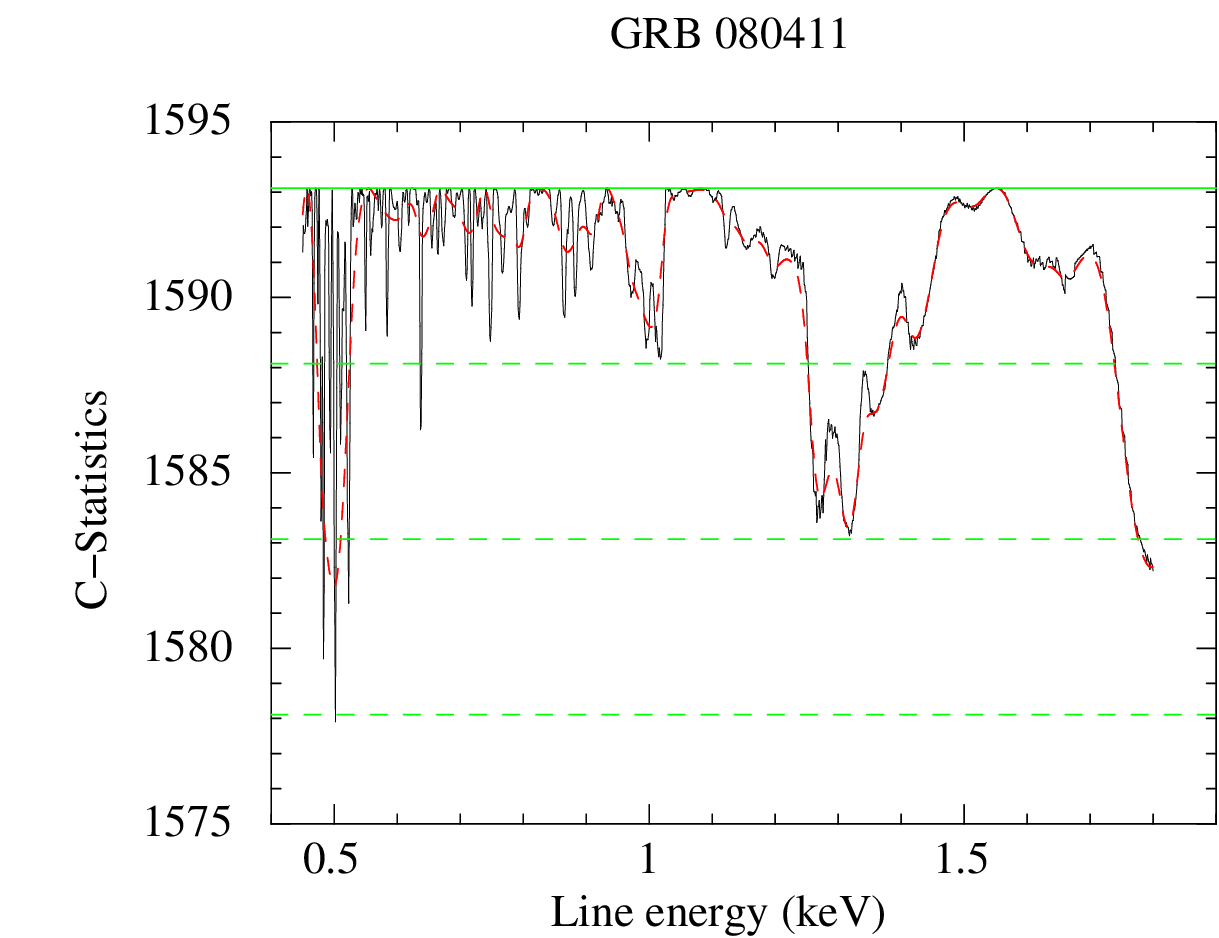}
\hspace{0.5cm}
\includegraphics[width=6cm,angle=-90]{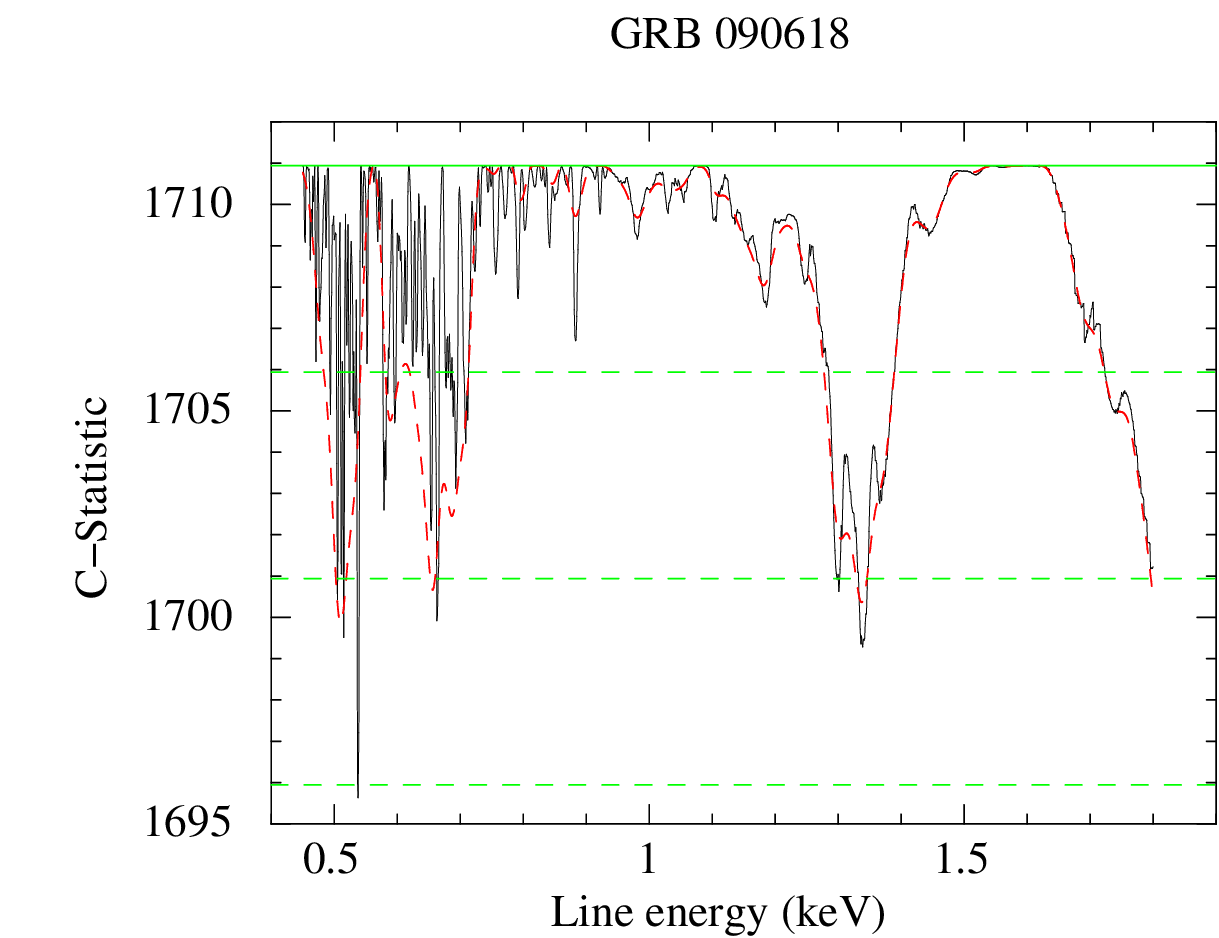}
}
\hbox{
\includegraphics[width=6cm,angle=-90]{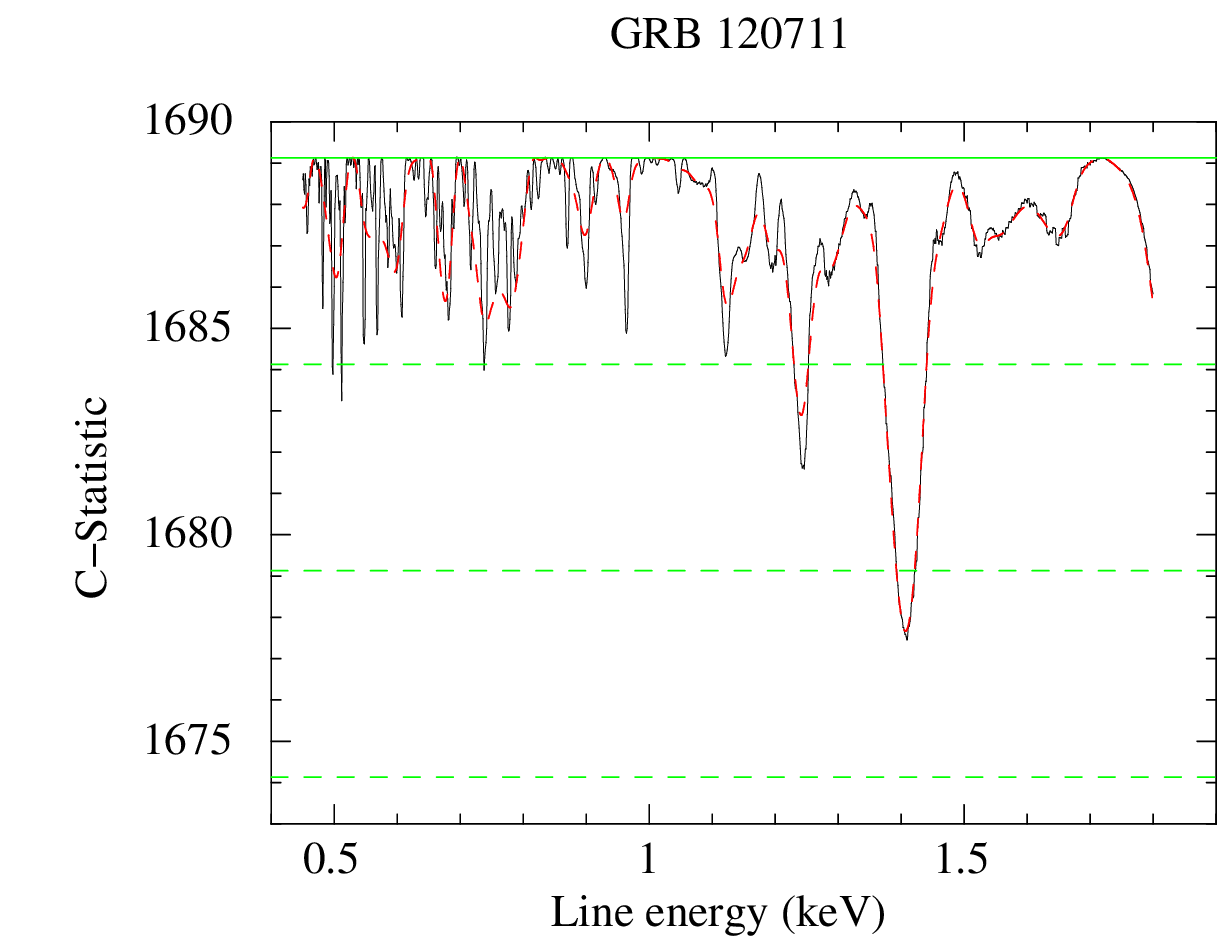}
\hspace{0.5cm}
\includegraphics[width=6cm,angle=-90]{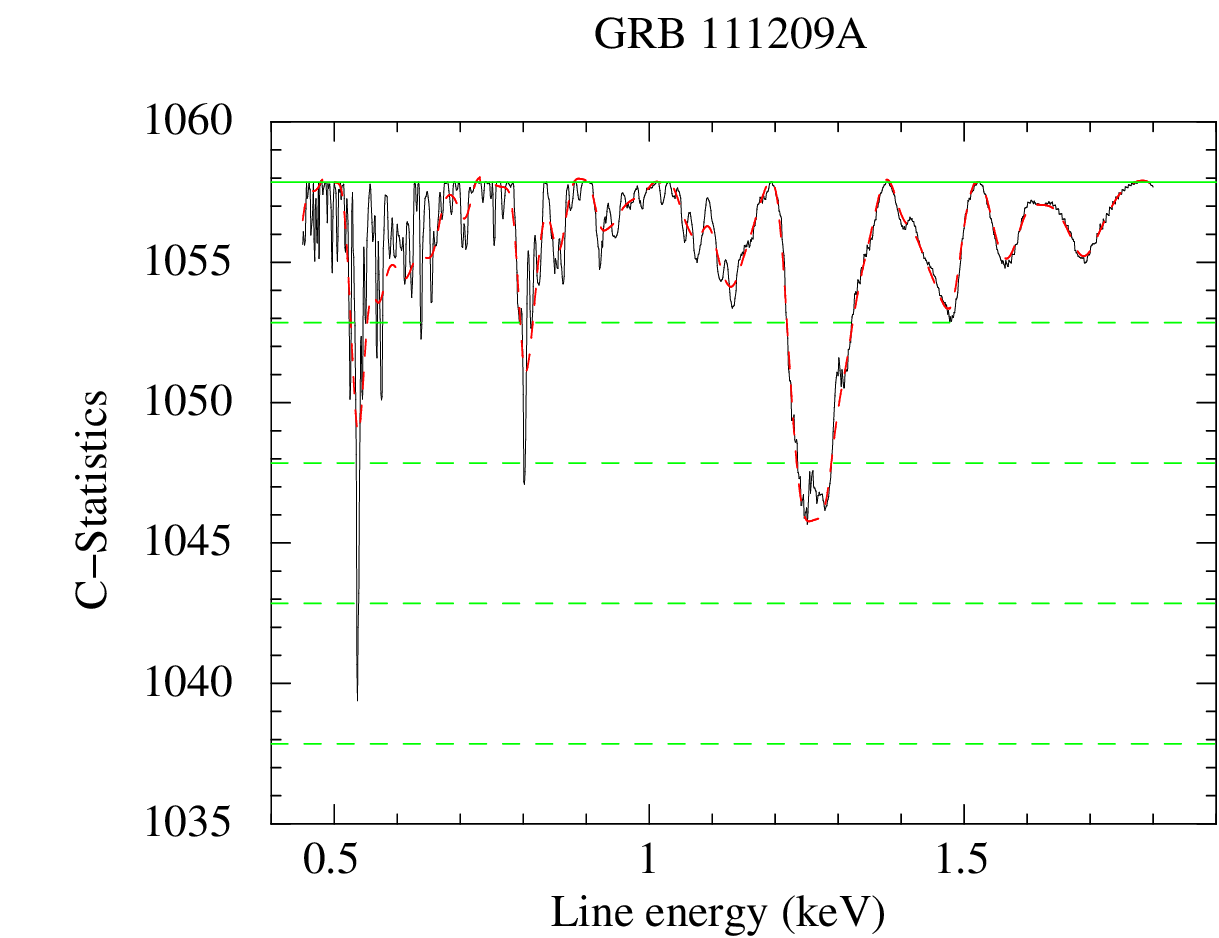}
}
\caption{Improvement in the C-statistic best-fit of each GRB as a result of including a narrow emission line in the 0.45--1.8 keV energy range.
The continuous (black) line refers to the inclusion of an emission
or absorption line with zero width, the dashed (red) line to a line with 10 eV width.
The solid horizontal line marks the best-fit C-statistic value without any additional lines. Dashed horizontal lines are drawn in 
steps of $\Delta{\rm C}=5$.}
\label{lines}
\vskip -0.1truecm
\end{figure*}

Emission lines in the X--ray spectra of GRBs have been discussed at length in the past years. 
The main emphasis was put on iron emission lines (having the largest yield). Neutral or ionised iron lines were 
reported in the spectra of GRB 970508 (BeppoSAX, Piro et al. 1999), GRB 970828 (ASCA, Yoshida et al. 2001), 
GRB 991216 (Chandra, Piro et al. 2000), and GRB 000214 (BeppoSAX, Antonelli et al. 2000). 
Detection of emission lines from elements at soft energies was reported for GRB 011211 (XMM-{\it Newton}, 
Reeves et al. 2002), GRB 030227 (XMM-{\it Newton}, Watson et al. 2003), and GRB 020813 (Chandra, Butler et al. 2003). 
All these lines were reported to be at a significance level higher than $3\,\sigma$ with a maximum of $4.7\,\sigma$ for 
GRB 991216, often estimated with an $F-$test. 
A thorough analysis assessing the statistical evidence for these line features was carried out by Sako, Harrison \& Rutledge 
(2005; see also Protassov et al. 2002). They selected spectra of sufficient statistical quality to allow for meaningful line searches 
(i.e. with more than 100 counts). They performed Monte Carlo simulations to search for discrete features. 
They found four emission features above a $3\,\sigma$ level that did not pass a further scrutiny.
They concluded that no credible X--ray line feature was detected. 

Emission lines have been searched for in spectra obtained with
the {\it Swift} (Gehrels et al. 2004) X--Ray Telescope (Burrows et al. 2005).
Moretti et al. (2008) performed a time-resolved spectral analysis of a bright GRB sample restricted to the 13 
softest spectra (power-law photon index $\Gamma>3$). Four of these GRBs show evidence of an additional (harder) component,
and in one case (GRB 060904B, {   see also Margutti et al. 2008}) this is consistent with a broad nickel emission line (significance $2.2-3.2\,\sigma$).
Hurkett et al. (2008) investigated 40 GRBs promptly observed by {\it Swift}/XRT and found no strong evidence of emission lines. 

In this paper we undertake a different approach. We search for narrow absorption (or emission) lines in the afterglow 
spectra of bright GRBs. These features arise as bound-bound transitions from metals and can be detected at soft X--ray energies. 
The strongest features that might be expected come from C, N, O, and Ne (i.e. the most abundant metals in a gas with a solar 
composition) and He- and H-like transitions. However, these lines depend on the radiation pattern (source luminosity 
and distance) and density, which leads to the exploration of a new frontier.
To this aim we need a large number of photons therefore we limit our analysis to the brightest GRBs with redshifts observed with the 
Reflection Grating Spectrometers (RGS; den Herder et al. 2001) on XMM-{\it Newton} (i.e. the largest throughput X--ray focusing facility). 

The paper is organised as follows. In Sect. 2 we describe the sample selection and the extraction of the data.
In Sect. 3 we fit the spectra and assess the statistical significance of the detected features.
In Sect. 4 we discuss our results and draw our conclusions.

\section{Sample selection and data extraction}

To search for narrow line features in the RGS spectra of GRBs, we need a large number of counts. For this reason we 
carried out our analysis of GRBs with known redshift and with a 0.3--3 keV fluence higher than $10^{-7}$ erg cm$^{-2}$. 
Our sample comprises five GRBs: 060729 ($z = 0.54$), 061121 ($z = 1.31$), 080411 ($z = 1.03$), 090618 ($z = 0.54$), and
GRB 120711A ($z=1.41$). We also included the very long burst GRB 111209A with a fluence lower by a factor of $\sim 2$ 
than our threshold because of the interest in it.
XMM-{\it Newton} data have already been analysed focusing on EPIC data: GRB 060729 (Grupe et al. 2007), GRB 061121 
(Page et al. 2007), GRB 090618 (Campana et al. 2011), and GRB 111209A (Gendre et al. 2013; Levan et al. 2013).
Data of GRB 080411 and GRB 120711A have not yet been analysed.
These GRBs are all bright and (relatively) close and were observed by XMM-{\it Newton} within one day from the prompt emission, 
guaranteeing a high observed fluence (see Table 1).

Data reduction was performed with the XMM-{\it Newton} Science Analysis Software (SAS) version xmmsas\_20131209\_1901-13.0.0 
and the latest calibration files. Data were locally reprocessed with  {\tt rgsproc}, {\tt emproc}, and {\tt epproc} . 
The RGS data were reduced using the standard SAS task {\tt rgsproc} and the most recent calibration files. 
For the high-background time filtering we  applied  a threshold of  0.2 counts s$^{-1}$  on the background event files.   
We then adopted a spectral binning  at half  the FWHM resolution of the instrument ($\Delta \lambda\sim 0.05\AA$),  and
 the C-statistic was employed in all the fits.
EPIC data were grade filtered using pattern 0--12 (0--4) for MOS (pn) data, and {\tt FLAG==0} and \#XMMEA\_EM(P) options.
RGS data were extracted from a standard region.
The pn and MOS events were extracted from a circular region of 870 pixels centred on source. 
Background events were extracted from similar regions close to the source and free of sources.

The RGS data were rebinned by ten original energy channels. MOS and pn data were rebinned to have 20 counts per energy bin.
RGS data were retained within the restricted 0.45--1.8 keV energy range (to have at least 25 cm$^2$ of effective area), 
MOS data were summed and fitted within the {  0.4--10} keV range, pn data within the {  0.35--10} keV range.

\begin{figure*}
\hbox{
\includegraphics[width=6cm,angle=-90]{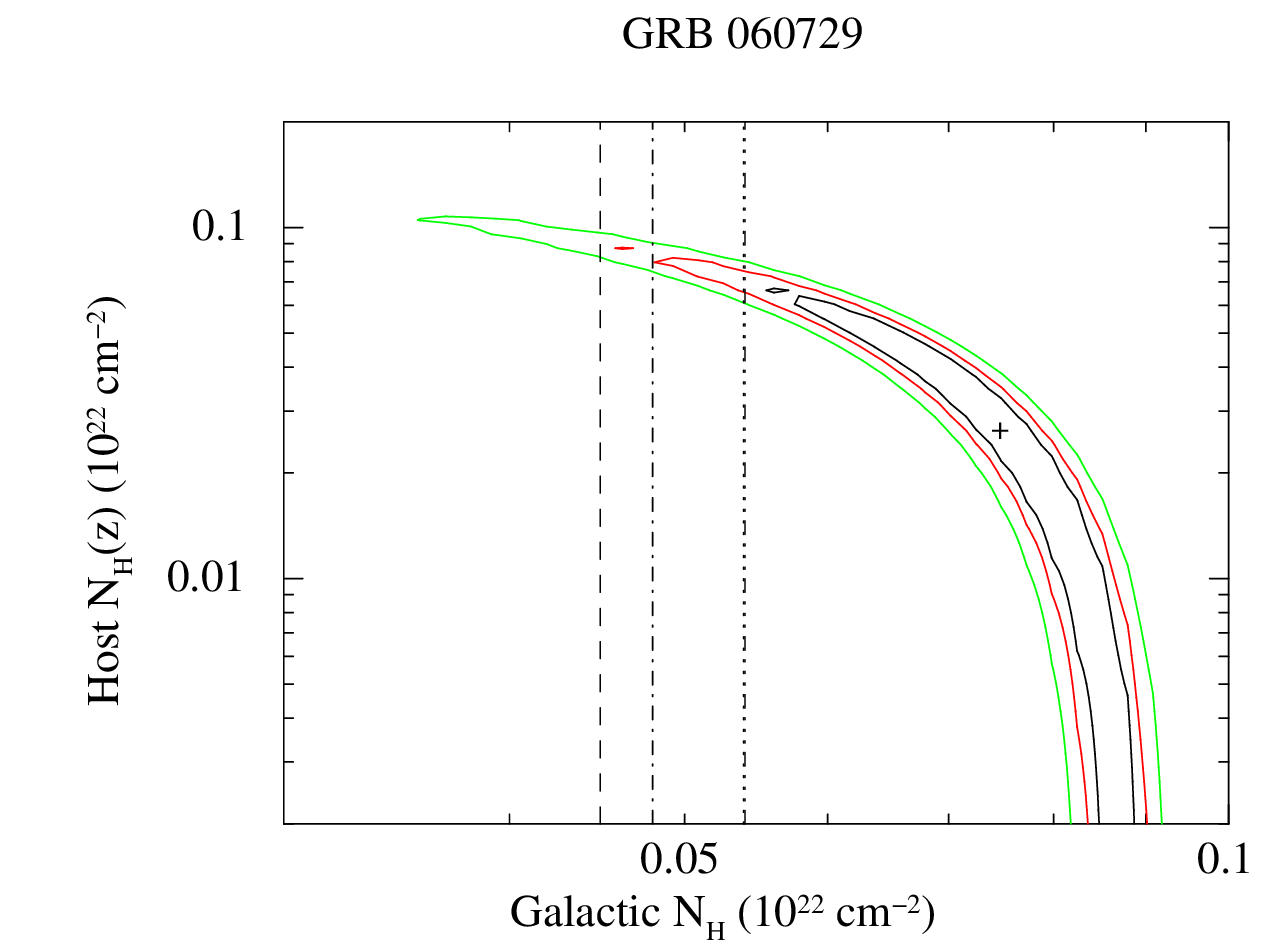}
\hspace{0.5cm}
\includegraphics[width=6cm,angle=-90]{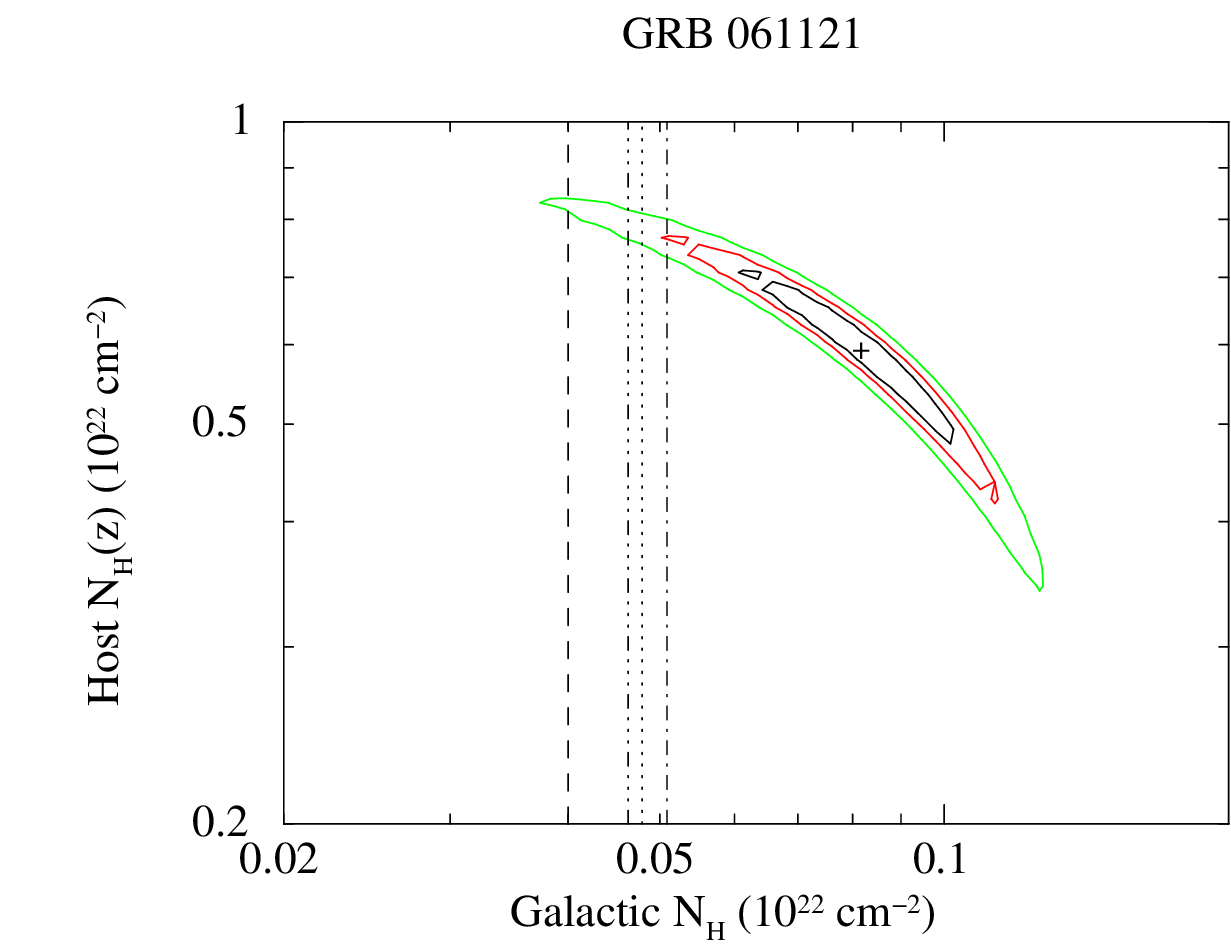}
}
\hbox{
\includegraphics[width=6cm,angle=-90]{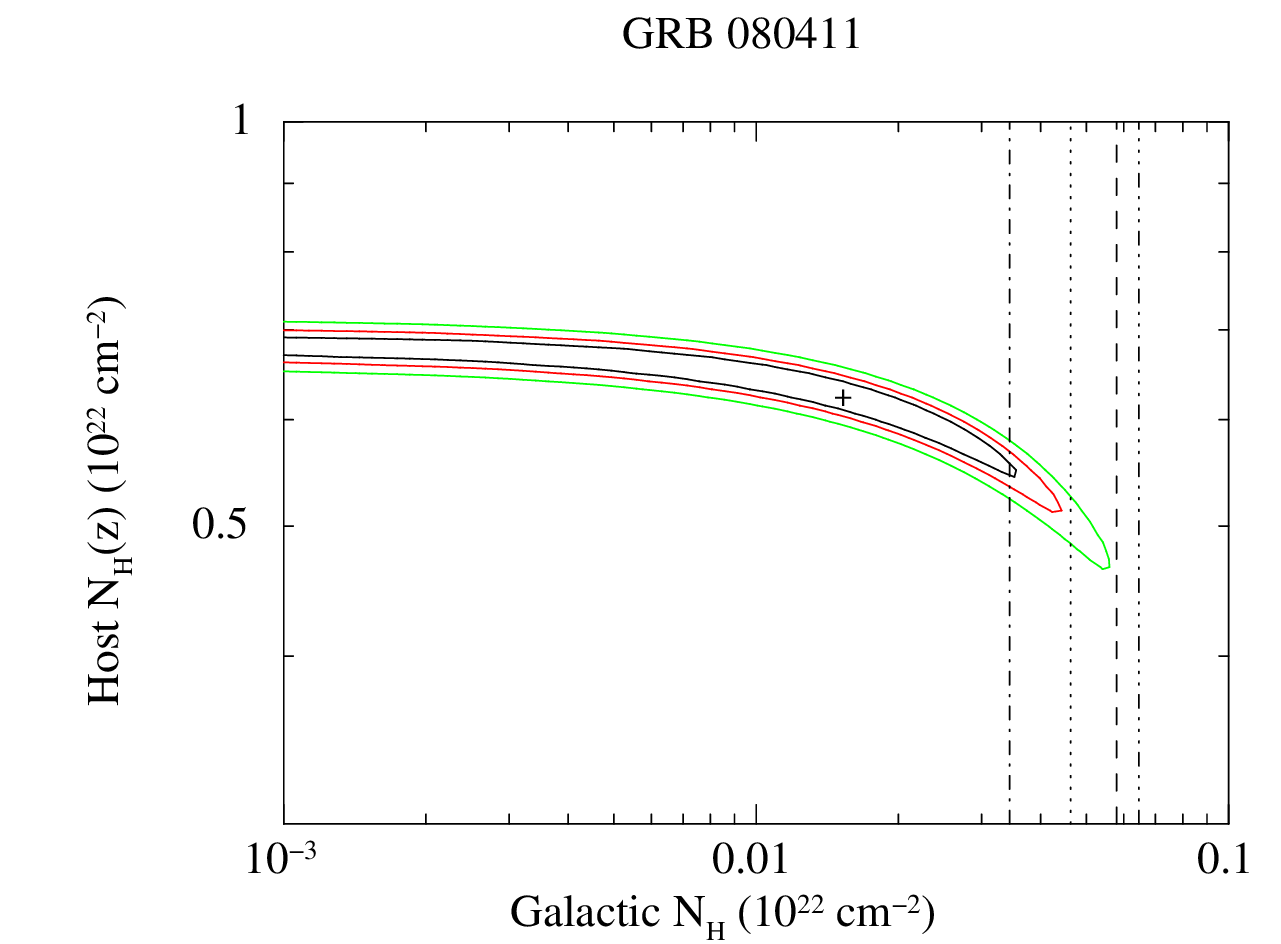}
\hspace{0.5cm}
\includegraphics[width=6cm,angle=-90]{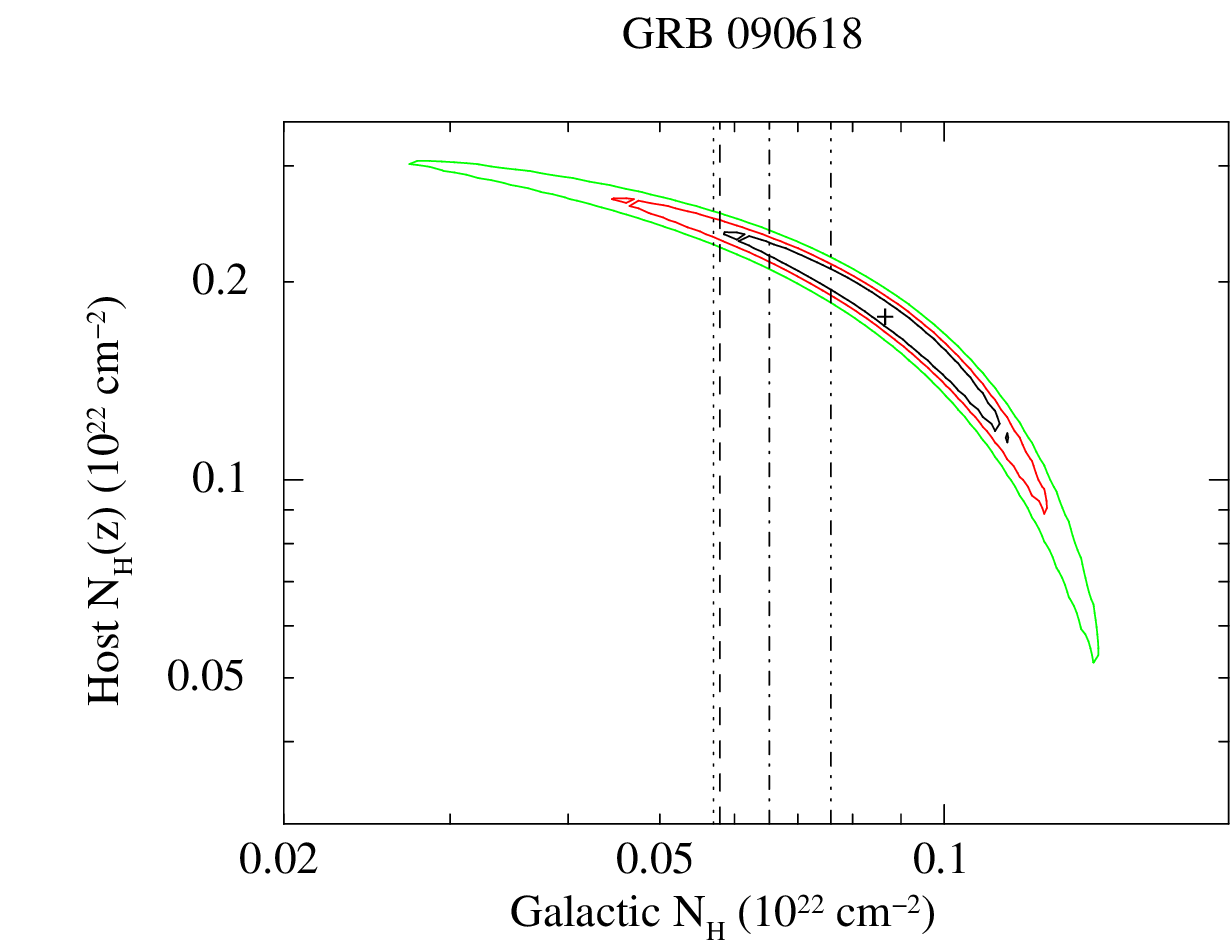}
}
\hbox{
\includegraphics[width=6cm,angle=-90]{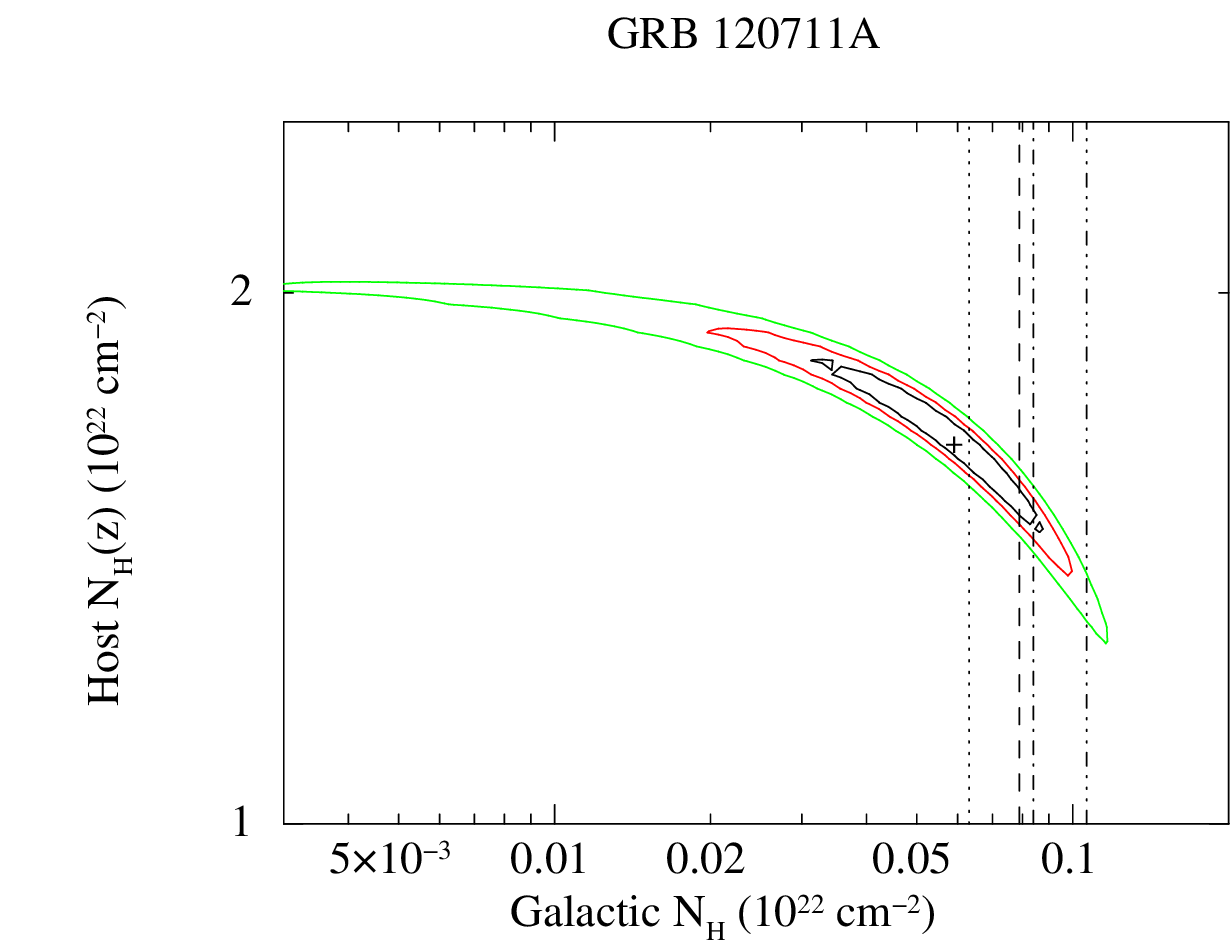}
\hspace{0.5cm}
\includegraphics[width=6cm,angle=-90]{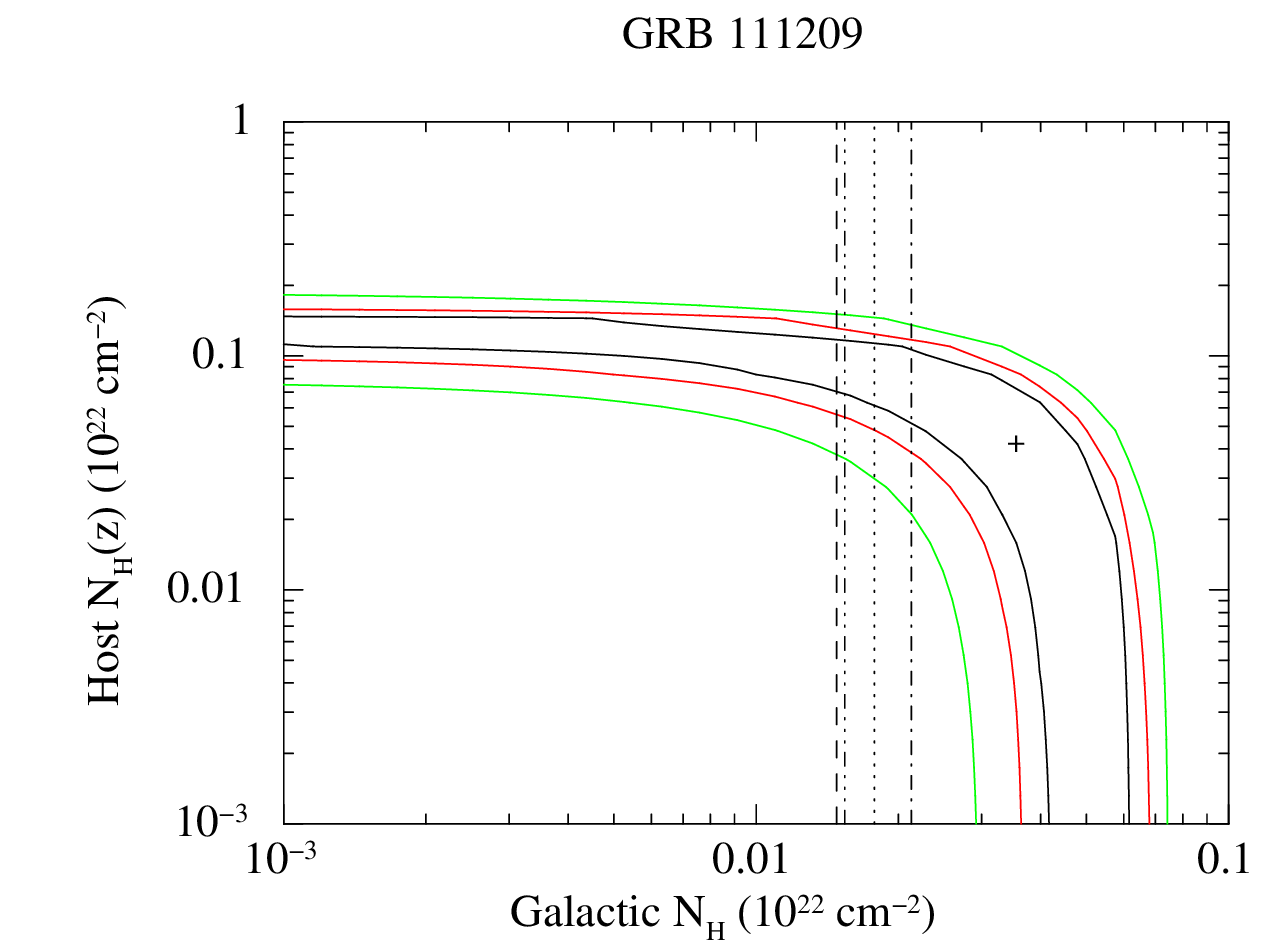}
}
\caption{Galactic vs. intrinsic column densities contour plot for the six bursts of our samples. Vertical lines mark the Galactic column density
derived from different estimates: dotted lines indicates the value adopted in the present paper (Kalberla et al. (2005), the
value includes 
a 10\% uncertainty); dashed lines the Kalberla et al. (2005) estimates, 
dot-dashed lines the Dickey \& Lockman (1990) estimates, and dot-dot-dashed lines the Willingale et al. (2013) estimates. 
The cross in each contour plot marks the best-fit value. In the first panel the dot-dot-dashed line and the dotted line coincide.}
\label{nhlines}
\vskip -0.1truecm
\end{figure*}

\section{Data analysis}

The GRB afterglow spectra are relatively simple, especially a few hours after the the main event.
We fitted all the spectra with an absorbed power-law model. The absorption pattern was composed of a Galactic absorption 
{  left free to vary within $\pm20\%$} of the Galactic value as derived in Kalberla et al. (2005). A second intrinsic component at the redshift of the 
GRB was included and free to vary. The absorption components were modelled with the {\tt TBABS} component in XSPEC
(Wilms, Allen \& McCray 2000) and using {\tt wilm} abundances and {\tt vern} absorption cross-sections within XSPEC.
Although the data were binned, C-statistics was used. 

We first fitted the RGS1, RGS2, pn, and MOS1+2 spectra with the absorbed power-law model, as above, including a normalisation
constant across the different instruments. Parameters are reported in Table \ref{fit}.
For GRB 111209A an additional emission component is needed to achieve a satisfactory fit. 
Following what has been found for other nearby GRBs showing a soft component (e.g. Campana et al. 2006b),
we added a black-body component.

\begin{table*}
\caption{GRB spectral fits.}
\label{fit}
\begin{center}
\begin{tabular}{cccccc}
\hline
GRB               &$N_H({\rm Gal})^*$      &Gal. latitude & $N_H(z)$                    & Power-law    & C-stat\\ 
                       &$10^{20}$ cm$^{-2}$&($b$ deg)      & $10^{21}$ cm$^{-2}$& photon index& (dof)\\
\hline
GRB 060729  & 5.39     (4.49)          & --27.3            &$0.7^{+0.1}_{-0.2}$  &$2.02^{+0.01}_{-0.01}$&1778.0 (1643)\\
GRB 061121  & 4.79     (4.00)          & +30.1            &$7.6^{+0.5}_{-0.2}$  &$1.95^{+0.01}_{-0.01}$&1412.8 (1449)\\
GRB 080411  & 4.63      (5.79)         & --43.6            &$5.1^{+0.3}_{-0.2}$  &$1.94^{+0.01}_{-0.01}$&1596.9 (1545) \\
GRB 090618  & 5.70     (5.79)         & +24.3             &$2.2^{+0.1}_{-0.1}$  &$1.90^{+0.01}_{-0.01}$&1709.6 (1652)\\
GRB 120711A& 6.31     (7.89)         & --28.2             &$16.3^{+1.9}_{-0.4}$&$1.89^{+0.01}_{-0.02}$&1689.1 (1616)\\
\hline
GRB 111209  & 1.78        (1.48)      & --70.3             &$0.9^{+0.1}_{-0.1}$  &$2.09^{+0.06}_{-0.06}$&1057.9 (993)\\
\hline
\end{tabular}
\end{center}
{\leftline{$^*$ Galactic column densities refer to the Kalberla et al. (2005) value. {  The quoted value is derived from the fit leaving the Galactic}}}
{\leftline{{  column density free to vary within an interval $\pm20\%$ of the Kalberla et al. (2005) value. The Kalberla et al. (2005) values are instead }}}
{\leftline{{  reported in parenthesis.}}}
{\leftline{Errors have been computed for $\Delta {\rm C}=2.71$ ($90\%$ confidence level for one parameter of interest).}}
{\leftline{For GRB 111209A we added a black-body component with $k\,T=0.22\pm0.01$ keV and
equivalent radius $R=(1.3\pm0.2)\times 10^{11}$ cm.}}

\end{table*}

\begin{table*}
\caption{Detected lines in GRBs ($\Delta {\rm C}>15$).}
\begin{center}
\begin{tabular}{cccccccc}
\hline
GRB               &Line energy                               & Line                    &Line energy     & Identification  --                & $\Delta$C& {  Single trial} signif. & {  Overall} signif.\\
                        &keV  (obs.) (E/A)                       &  width (eV)         & rest keV (\AA)&  energy (keV)                    &                    &(Gauss. $\sigma$) & (Gauss. $\sigma$) \\
\hline
GRB 060729&$0.524^{+0.003}_{-0.003}$ E& --                          &  --                  & O K$\alpha$ Gal. -- 0. 524 & 16.0          & 2.85 & 2.22\\
GRB 060729&$0.500^{+0.004}_{-0.004}$ E&$5.0^{+2}_{-2}$  & 0.770 (16.1)& Co L$\alpha$ -- 0.776        & 17.0          & 2.96 & 2.36\\
GRB 061121&$0.483^{+0.013}_{-0.015}$ E&$14^{+14}_{-9}$& 1.118 (11.1)& ??                                           & 15.7           & 2.69 & 2.02\\
GRB 090618&$0.538^{+0.001}_{-0.001}$ E& --                          & --                   & O edge Gal. -- 0.543            & 15.2           & 2.46 & 1.76\\
GRB 111209&$0.537^{+0.001}_{-0.001}$ E& --                          & --                   & O edge Gal. -- 0.543            & 18.9           & 3.50 & 3.00 \\
\hline
\end{tabular}
\end{center}
\label{tablines}
\end{table*}

We then included a line, either in emission or in absorption, in the fit.
To assess its presence, we fixed its width (either to zero, i.e. instrumental, or to 10 eV, about twice the spectral resolution).
We probed the significance of the line by evaluating the change in C-statistics when the energy varied 
in steps of 1 eV in the 0.45--1.8 keV energy range of the RGS instruments. 
At a fixed energy and width, the only parameter to vary is the line normalisation.
In Fig. \ref {lines} we report the C-statistics variations induced by including a fixed energy line
with respect to the best-fit value with no line included.
This has to be intended as an initial explorative search for lines. Only tentative lines producing $\Delta {\rm C}>10$ were 
taken into further account.

These tentative lines were further investigated including in the model a Gaussian with fixed (zero) or free width, using as a starting 
point the line energy derived from the line search above.
After this step we retained only lines producing $\Delta {\rm C}>15$. 

We tentatively detected five lines (all in emission) in the GRB spectra under consideration. Three of them are clearly related to oxygen in our Galaxy, 
either to the K$\alpha$ line or to the absorption edge (see Table \ref{tablines}).
The line in GRB 060729 might be consistent with a cobalt L$\alpha$ line. The line in GRB 061121 did not match any template energy. 
We also tried to leave free the abundances of the elements whose edges fall within the RGS energy range (i.e. O, F, Ne, Na, Mg, and Al, using 
{\tt TBVARABS}) keeping {  this time} the Galactic column density fixed to the Kalberla et al. value. All the abundances remained undetermined except for
oxygen, which was found to be consistent with the solar abundance value.

\subsection{Assessing the statistical significance}

To asses the significance of the line detections, we  performed extensive Monte Carlo simulations. 
For each observation we assumed as our null hypothesis model the best-fit model for the combined fit of all the 
XMM-{\it Newton}  RGS and EPIC spectra, with no emission or absorption lines. 
We then simulated for each GRB 1,000  spectra for each camera similar to the available observations. 
Each set of simulated spectra was fitted with the null hypothesis model to obtain the C value. 
We then searched for an emission or absorption Gaussian line across the same energy range that  was adopted for the spectral fitting. 
The Gaussian line was assumed to be unresolved at the RGS resolution, and we stepped the energy centroid in increments of 10 eV,  
refitting at each step. 
We recorded  the  lowest
C value for each of the simulated set of spectra and constructed   a distribution 
of the $\Delta C$ with respect to the null hypothesis model for each GRB; from this distribution we then determined the  $\Delta C$ that would 
correspond to a detection significant at $>99.90\%$. 
Line significance is reported in Table \ref{tablines} and ranges between 2.5 and 3.5$\,\sigma$ for one single GRB at a time
(see Table \ref{tablines}).

\begin{table*}
\caption{Neon and oxygen absorption lines search.}
\begin{center}
\begin{tabular}{ccccc}
\hline
GRB                 &Ne IX (0.922 keV)& Ne X (1.021 keV) & OVII (0.568 keV) & OVIII (0.654 keV) \\
                          &EW (eV)                 & EW (eV)                 & EW (eV)               & EW (eV)                \\ 
\hline
GRB 060729  &$<2.0$                    & $<1.2$                    & $<15.1$              & $<7.7$                    \\ 
GRB 061121  &$<23.6$                  & $<4.4$                    & --                           & $<145$                   \\ 
GRB 080411  &$<14.6$                  & $<6.1$                    & $<30.8$              & $<103$                    \\ 
GRB 090618  &$<5.2$                    & $<5.1$                    & $<21.1$              & $<8.7$                      \\ 
GRB 120711A&$<5.4$                    & $<2.4$                    & --                           & --                              \\ 
GRB 111209  &$<1.0$                    & $<1.6$                    & $<37.6$               & $<16.4$                    \\
\hline
\end{tabular}
\end{center}
\label{narrow}
\end{table*}

\subsection{Narrow absorption line search}

In addition to the blind search for narrow features described above, we also searched for selected absorption lines in the XMM-{\it Newton} spectra.
We selected four prominent X--ray lines (Ne IX, Ne X, O VII, and OVIII) and included an absorption line in the X--ray spectra, one at a time.
The line was modelled as a Gaussian with zero-width (i.e. spread only by instrumental resolution) at the GRB redshift with a negative normalisation (i.e. 
an absorption line). We extended the pn energy range down to 0.2 keV and the MOS to 0.3 keV to derive limits for the GRBs at a redshift higher than 1.
No lines were detected. We then worked out the $90\%$ limit for the normalisation, fixed the normalisation to this upper limit, refitted the spectrum, and derived the 
$90\%$ confidence limit on the line equivalent width. These limits are reported in Table \ref{narrow}. Especially when the lines fall within the RGS energy range,
limits are very tight, at the level of a few eV. Looser limits are derived for oxygen lines because they are probed at a poorer energy resolution.

%
%

\subsection{Hot medium search}

In addition to the cold absorbing medium within the GRB host galaxy, we searched for the presence of a hot medium.
This was done by adding a hot component to our best-fit model, modelled with {\tt zxipcf} (Reeves et al. 2008). 
The hot absorber is not needed in any of the cases, and we derived limits on the ionisation parameter ($\xi=L/(n\,r^2)$, with $L$
the photoionising luminosity, $n$ the density medium, and $r$ the medium distance) and the corresponding absorbing column density
(we also assumed a complete covering of the absorbing medium).
These limits allow only for a heavily  ionised medium $\log\xi\gsim 3$ with a column density comparable to the cold one 
($N_H(z)\lsim 10^{21}\cmdue$, see Table \ref{hotm}). Clearly, the allowed column density 
increases for a very large ionisation parameter ($\log\xi=6$) because its effective contribution to the absorption is much lower (see Table \ref{hotm})\footnote{We note that this absorption model 
for low values of the ionisation parameter $\log\xi \lsim -1$ provides incorrect results and is unreliable
.}.

\begin{table}
\caption{Limits on the hot absorbing medium.}
\begin{center}
\begin{tabular}{c|cc|cc}
\hline
GRB                 &$\log\xi$   & $N_H(z)$                 &  $\log\xi$    & $N_H(z)$                      \\
                          &                  & ($10^{21}\cmdue$)&                     & ($10^{23}\cmdue)$        \\ 
\hline
GRB 060729  &$>3.1$     &$<0.7$                       & 6                  & $<4.5$   \\
GRB 061121  &$>2.9$     &$<1.0$                       & 6                  & $<30$   \\   
GRB 080411  &$>3.0$     &$<0.9$                       & 6                  & $<8.5$   \\
GRB 090618  &$2.7^{+0.6}_{-0.3}$ &$<3.7$       & --                  & --   \\
GRB 120711A&$>3.6$     &$<12$                        & 6                  & $<25$   \\
GRB 111209  &$>3.2$     &$<1.1$                       & 6                  & $<35$   \\
\hline
\end{tabular}
\end{center}
\label{hotm}
\end{table}

\subsection{Column densities}

Given the quality of our data, we investigated  the absorption component of the GRBs in our sampling in greater detail.
This is also motivated by our finding lines close to the expected energy of the Galactic oxygen edge.
We fitted the data by leaving the Galactic and extragalactic absorption components free.
Contour plots of the Galactic vs.  intrinsic column densities are shown in Fig. 
\ref{nhlines}. We then compared these values with those that can be found in the literature: Kalberla et al. (2005), 
Dickey \& Lockman (1990), both based on radio data, and Willingale et al. (2013), starting with the Kalberla et al. (2005) 
data and correcting them for the presence of H$_2$.
In four cases (out of six) these tabulated estimates fall below the best-fit value. This suggests that in these four GRBs, 
and only in these GRBs, we find a hint of Galactic absorption in addition to the tabulated Galactic value.
In the line searches we fixed the Galactic column density to the Kalberla value. This missing absorption component was 
then compensated for in the fit by adding a line. 
This would indicate a Galactic origin of the (low-significance) features we identified and would probably identify them as due to 
the presence of more oxygen along the line of sight with respect to the tabulated Galactic column densities.
Although our sample is small, we derived the ratio of the best Galactic $N_H$ with respect to the tabulated values.
We worked out a $\chi^2$ based on the best-fit column density and its ($1\,\sigma$) error compared with the tabulated values.
We obtained a reduced $\chi^2_{\rm red}=1.16$, 0.85 and 1.15 (with 6 degrees of freedom) for the values of Kalberla et al. (2005), 
Dickey \& Lockman (1990), and Willingale et al. (2013), respectively. Null hypothesis probabilities are 33\%, 53\%, and 33\%, 
respectively. This indicates that the Dickey \& Lockman (1990) Galactic column densities fitted with the {\tt tbabs} model provide a 
slightly better description of the six randomly sampled ($|b|\gsim 25$\deg) lines of sight.

\subsection{Helium absorber}

Watson et al. (2013) envisaged an He absorber as the main contributor to absorption in the host galaxy.
This idea relies on possible ionisation of the star-forming region where the GRB explodes through the massive stars formed in it. This will give rise to an HII
region. Since hydrogen is fully ionised, the dominant absorber in the X--ray is represented by helium because of its higher ionisation energy (54 eV).
In addition, the GRB progenitor itself can expel He shells during its lifetime
to give rise to a Wolf-Rayet star, which is thought to be the 
progenitor of type Ic supernovae. These He shells might contribute by absorbing GRB photons. 
Here we investigate the possible role of a He absorber. We fitted the spectra using {\tt tbvarabs} within XSPEC, allowing only for a helium 
(and hydrogen) column density, but setting the column densities of all the other elements to zero.
The fits are almost as good as those obtained for a cold absorber (see Table \ref {he}).
An overall fit, grouping all the GRB data together, provides a null hypothesis probability of $0.5\%$ (reduced $\chi^2=1.039$ for 8897 dof) 
for the cold absorber model and of $0.03\%$  (reduced $\chi^2=1.051$) for the helium model, suggesting a slight preference for 
cold absorber.

\begin{table}
\caption{GRB spectral fits with a helium absorber.}
\label{he}
\begin{center}
\begin{tabular}{cccc}
\hline
GRB                  & $N_H(z)^{\rm He}$      &  C-stat               & C-stat\\ 
                          & $10^{21}$ cm$^{-2}$   &  (dof)                  & cold absorb.\\
\hline
GRB 060729  &$1.5^{+0.2}_{-0.1}$        & 1786.4 (1643) & 1778.0\\
GRB 061121  &$25.6^{+0.9}_{-0.9}$      & 1443.8 (1449) &1412.8 \\
GRB 080411  &$24.1^{+3.2}_{-1.9}$      & 1568.2 (1545) &1596.9 \\
GRB 090618  &$5.0^{+0.2}_{-0.2}$        & 1777.2 (1651) &1709.6 \\
GRB 120711A&$54.1^{+1.6}_{-1.6}$     & 1704.7 (1616) &1689.1 \\
\hline
GRB 111209  &$4.3^{+4.3}_{-3.2}$        & 1072.6 (993)   &1057.9\\
\hline
\end{tabular}
\end{center}
\end{table}

\section{Discussion and conclusions}

We carried out a search for narrow emission or absorption lines in the RGS (and EPIC) spectra of bright GRB afterglows observed
by XMM-{\it Newton}. GRBs were selected to have an observed fluence $>10^{-7}$ erg cm$^{-2}$ to provide 
$\gsim 5,000$ counts in the RGS. In addition to the intrinsic absorption typically observed in the X--ray spectra of GRB afterglows,
we searched for narrow absorption or emission line features that
are possibly related to a dense medium close to the GRB site.
We detected some (emission) line features in the  XMM-{\it Newton} spectra. A deep scrutiny with Monte Carlo simulations showed that these lines
are at a significance of $2.5-3.5\,\sigma$ (for each single GRB). 
Considering the number of trials (6), the highest significance feature ($3.5\,\sigma$) reduces its significance to $3.0\,\sigma$ (see also Table 3).
Thus, only the emission feature in GRB 111209 survives a deep statistical scrutiny (i.e. significance higher than $3\,\sigma$). 
However, the line energy of this  feature is consistent with the (Galactic) oxygen edge (i.e. 0.5 keV). 
The significance was evaluated keeping the Galactic column density within $\pm20\%$ of the Kalberla et al. value.  
When we left the Galactic column density free to vary, this emission feature 
disappeared and, consequently, the value of the Galactic column density increased. This might indicate that the values of the Galactic column densities 
from wide-field HI radio maps are underestimated by up to a factor of $\sim 2$. In addition, there is the long-standing problem of oxygen 
abundance because oxygen is the main contributor to the total column density. The value reported in Anders \& Grevesse (1989) and 
Grevesse \& Sauval (1998) is a factor of two higher than the one adopted by Wilms, et al. (2000) and Asplund et al. (2009).
Recent studies seem to support the Asplund et al. findings, however (Gatuzz et al. 2014; Baumgartner \& Mushotzky 2006).
Summarising, we consider the features at $\sim 0.5$ keV  to be
Galactic in origin and related to a slight increase of the column density with respect to 
HI maps or and slight overabundance of oxygen along the line of sight.
The remaining two features are more puzzling. They are not connected to oxygen. One lies close to the expected energy of the cobalt 
K$\alpha$ line (GRB 060729), whereas the other is not easily identifiable with the main emission lines, even moderately 
redshifted or blueshifted. However, their overall statistical significance is low and we do not consider them as real.
Based on this analysis, we can place a firm upper limit of $\sim 10$ eV on the equivalent width of narrow features in the GRB 
afterglow spectra of our sample in the RGS energy range, and a logarithmic average of the features that did not pass our tests of $5$ eV.

Searches for emission lines in the X--ray afterglow spectra have been carried out in the past. Hurkett et al. (2008) searched for lines in 
40 {\it Swift} GRBs within a few ks from the explosion. No lines were detected. The logarithmic mean equivalent width of their possible features 
was $\sim 80$ eV.
A similar search was performed by Butler (2007) on 70 {\it Swift} GRB afterglow spectra. The outcome of this investigation was not conclusive. 
The logarithmic mean equivalent width of the detected features is $\sim 40$ eV. 

A further complication might arise from the presence of a soft component in the afterglow X--ray spectra. This component has been detected in 
nearly $50\%$ of close ($z<1.5$) GRBs (Friis \& Watson 2013; Sparre \& Starling 2012). 
This component is present at early times (typically shorter than a few hours), which means 
that it should not be present in the XMM-{\it Newton} spectra we investigates. This is not the case of the ultra-long GRB 111209, where a soft 
black-body component has been detected  (Gendre et al. 2013; Levan et al. 2013), which we included in our fit.  

The most striking result is the absence of narrow absorption features. In all the GRBs of our sample there is the need for an intrinsic absorber 
at the GRB redshift, and we expect to see the imprint of this medium on the X--ray spectrum in addition to the cumulative effect of the column density.
However, an unsupervised search as well as a search for neon or oxygen high-ionisation lines was negative. We also searched for a hot medium with negative results. One possible solution is that absorption in the host galaxy mainly comes from a large amount of helium in the natal HII 
region that is undetected at UV/optical wavelengths (Watson et al. 2013). A spectrum with no features imprinted by the absorber is expected in this case, 
as observed. It remains to understand, however, how such a large amount of helium 
can be present and not have been photoionised by the extremely high photon flux of the GRB itself.
In addition, the overall fit of all the GRB spectra indicates a mild preference for the cold absorber case. Based on spectral fits, the cold absorber 
case provides an overall null hypothesis probability of $0.5\%$, whereas the fit with a helium absorber has a probability of only $0.03\%$. A warm absorber 
(instead of a hot one) might provide an alternative to be investigated in more detail (Prochaska et al. 2008; Schady et al. 2001;
Campana et al. 2016, in preparation).
%
%
Additional observations with high signal-to-noise ratios (e.g. in the near future with Athena) will help 
improve the description of the X--ray absorption from our Galaxy{  and in the GRB host galaxies}.

\section{Acknowledgments}
This work was partially supported by ASI contract I/004/11/1. We thank the referee, Darach Watson, for useful comments.

\end{document}